\documentclass[useAMS,usenatbib]{mn2e}
\usepackage{times}
\usepackage{arydshln}
\usepackage{color}

\usepackage[T1]{fontenc}

\usepackage{graphicx}	
\usepackage{amsmath}	
\usepackage{amssymb}	

\def\nustar{\textit{NuSTAR}}
\def\xmm{\textit{XMM-Newton}}
\def\xspec{\textsc{xspec}}
\def\rxte{\textit{RXTE}}

\title[Reflection tomography of H 1743--322]{Tomographic reflection
  modelling of quasi-periodic oscillations in the black hole binary H 1743--322}

\author[A. Ingram et al]{Adam
Ingram,$^{1}\thanks{E-mail:a.r.ingram@uva.nl}$
Michiel van der Klis,$^1$
Matthew Middleton,$^2$
\newauthor
Diego Altamirano,$^3$
\& Phil Uttley$^1$ \\
$^1$Anton Pannekoek Institute for Astronomy, University of Amsterdam,
Science Park 904, 1098 XH, Amsterdam, The Netherlands.\\
$^2$Institute of Astronomy, Cambridge University, Madingley Road, CB3 0HA, Cambridge, UK \\
$^3$Department of Physics \& Astronomy, University of Southampton,
Southampton, Hampshire SO17 1BJ, UK }

\date{Accepted 2016 October 4. Received 2016 September 25; in original
  form 2016 July 29}

\pubyear{2016}

\begin{document}
\label{firstpage}
\pagerange{\pageref{firstpage}--\pageref{lastpage}}
\maketitle

\begin{abstract}
Accreting stellar mass black holes (BHs) routinely exhibit Type-C
quasi-periodic oscillations (QPOs). These are often interpreted as
Lense-Thirring precession of the inner accretion flow, a relativistic
effect whereby the spin of the BH distorts the surrounding space-time,
inducing nodal precession. The best evidence for the precession model
is the recent discovery, using a long joint \xmm~and
\nustar~observation of H 1743--322, that the centroid energy of the
iron fluorescence line changes systematically with QPO phase. This was
interpreted as the inner flow illuminating different azimuths of the
accretion disc as it precesses, giving rise to a blue/red shifted iron
line when the approaching/receding disc material is illuminated. Here,
we develop a physical model for this interpretation, including a
self-consistent reflection continuum, and fit this to the same H
1743--322 data. We use an analytic function to parameterise the
asymmetric illumination pattern on the disc surface that would result
from inner flow precession, and find that the data are well described
if two bright patches rotate about the disc surface. This model is
preferred to alternatives considering an oscillating disc ionisation
parameter, disc inner radius and radial emissivity profile. We find
that the reflection fraction varies with QPO phase ($3.5\sigma$),
adding to the now formidable body of evidence that Type-C QPOs are a
geometric effect. This is the first example of tomographic QPO
modelling, initiating a powerful new technique that utilizes QPOs in
order to map the dynamics of accreting material close to the BH.
\end{abstract}




\section{Introduction}
\label{sec:intro}

In black hole (BH) X-ray binary systems, matter is accreted from the
binary partner through a geometrically thin, optically thick disc,
which emits a multi-coloured blackbody spectrum
(\citealt{Shakura1973}; \citealt{Novikov1973}). Compton up-scattering
of soft seed photons by a cloud of hot electrons close to the black
hole also contributes a power-law component to the X-ray spectrum,
with low and high energy cut-offs determined respectively by the seed
photon and electron temperatures (\citealt{Thorne1975};
\citealt{Sunyaev1979}). Some fraction of the Comptonized photons
reflect from the disc and are scattered into our line-of-sight. This
imprints characteristic reflection features onto the spectrum,
including a prominent iron K$_\alpha$ fluorescence line at $\sim 6.4$
keV and a so-called reflection hump, resulting from inelastic
free-electron scattering, peaking at $\sim 30$ keV
(\citealt{Ross2005}; \citealt{Garcia2013}). These reflection features
provide the opportunity to probe the dynamics of the accretion disc,
since the iron line is observed to be distorted by Doppler shifts from
orbital motion and gravitational redshift (\citealt{Fabian1989}).

So-called Type-C quasi-periodic oscillations (QPOs) are routinely
observed in the X-ray flux, with the oscillation frequency increasing
from $\sim 0.1-30$ Hz as the spectrum transitions from the hard
power-law dominated \textit{hard state} to the disc dominated \textit{soft state}
(e.g. \citealt{Wijnands1999}). In the \textit{truncated disc model},
the disc evaporates inside of some transition radius to form a
power-law emitting hot inner flow (\citealt{Ichimaru1977};
\citealt{Done2007}). The spectral transitions then arise as the disc
inner radius moves inwards, until it reaches the innermost stable
circular orbit (ISCO) in the soft state. Alternatives include a corona
partially covering the disc, confined by magnetic fields
(\citealt*{Galeev1979}; \citealt{Haardt1991}) and an outflowing jet
(\citealt{Markoff2005}). In all models, changes to the accretion
geometry are required to explain the spectral transitions.

Suggested QPO mechanisms in the literature either consider
instabilities in the accretion flow (e.g.\citealt{Tagger1999};
\citealt{Cabanac2010}), or a geometric oscillation
(e.g. \citealt{Stella1998}; \citealt{Wagoner2001}). There is now
strong evidence in favour of the geometric models, since high
inclination (more edge-on) systems display stronger QPOs than low
inclination (more face-on) systems (\citealt{Schnittman2006};
\citealt{Motta2015};
\citealt{Heil2015}). Phase lags between energy bands also strongly
depend on inclination, with hard photons lagging soft for low
inclination objects, and vice-versa for high inclination objects (van
den Eijnden et al, submitted). A prominent interpretation associates the
QPO with Lense-Thirring precession (\citealt{Stella1998}). This is a
General Relativistic effect whereby the spin of the BH induces
precession in orbits of particles inclined relative to the equatorial plane
(\citealt{Lense1918}). \cite*{Schnittman2006} considered a precessing
ring at the inner edge of the disc. However, the disc is expected to
be held stationary by viscosity (\citealt{Bardeen1975}), and the QPO
amplitude is stronger in the Comptonized spectrum than in the disc
spectrum (\citealt{Sobolewska2006};
\citealt{Axelsson2014}). \cite*{Ingram2009} instead suggested that the
entire inner flow precesses whilst the disc remains stationary,
motivated by the simulations of \cite{Fragile2007}.

\cite{Ingram2012a} showed that precession of the inner flow will cause
the iron line to rock from blue to red shifted, as the inner flow illuminates the
approaching followed by receding disc material. This prediction can be
directly tested with QPO phase-resolved spectroscopy. Phase-resolving
poses a technical challenge, since the stochastic nature of the QPO
prevents phase folding - as can be used for e.g. neutron star pulses
(e.g. \citealt{Wilkinson2011}; \citealt{Gierlinski2002}) - from being
appropriate. \cite{Miller2005} applied a simple flux selection to
strong QPOs from GRS 1915+105, but constraining spectra for more than
two phases requires a more sophisticated method. \citet[hereafter
IK15]{Ingram2015} developed a technique that uses the average Fourier
properties in order to reconstruct phase-resolved spectra, and used it
to discover spectral pivoting in GRS 1915+105. \cite{Stevens2016} used
a similarly sophisticated technique in order to measure changes in the
disc temperature of GX 339-4 during the QPO cycle (although this is
for a Type B QPO; see e.g. \citealt{Casella2005} for QPO
classifications). However, the phase-resolved behaviour of the iron 
line could not be constrained in these studies due to limitations on
data quality. \citet[hereafter I16]{Ingram2016} applied a
developed version of the IK15 method to a long \xmm~and
\nustar~observation of H 1743--322 in the hard state in order to
measure a QPO phase dependence of the iron line centroid energy. This
provides strong evidence that the QPO is driven by precession - either
precession of the inner flow (\citealt{Ingram2012a}), or alternatively
of the disc (i.e. precession of the reflector; \citealt{Schnittman2006}).

In this paper, we develop a physical model for the QPO phase-resolved
spectra measured in I16. Our model is designed to mimic illumination
of the disc by a precessing inner flow - as the flow precesses, it
preferentially illuminates different azimuths of the disc. Rather
than consider a specific geometry for the precessing flow, we
parameterise the asymmetric, rotating illumination profile with an
analytic function. This has the advantage of making no \textit{a
  priori} assumptions about the inner flow geometry, it enables the
model to be fast enough to fit directly to data, and it allows us to
define asymmetry parameters that can be set to zero in order to
recover the usual case of axisymmetric illumination. In Section
\ref{sec:data} we summarise the observations and the phase-resolving
method. In Section \ref{sec:cal}, we fit the time-averaged spectrum
with a relativistic reflection model in order to address
cross-calibration issues between \xmm~and \nustar, and also to provide
a comparison to our eventual tomographic modelling. We present the
details of our model in Section \ref{sec:model} and the results of our
tomographic modelling in Section \ref{sec:results}. We discuss our
results in Section \ref{sec:discussion} and present our conclusions in
Section \ref{sec:conclusions}.

\section{Data analysis}
\label{sec:data}

\subsection{Observations}

We consider \xmm~and \nustar~data from the 2014 outburst of H
1743--322. \xmm~observed this outburst for two full orbits around the
Earth in late September 2014. We use data from the EPIC pn, which was
in timing mode for the entire exposure. The first orbit has obs ID
0724400501, and the second orbit is split into two obs IDs (0724401901
and 0740980201) due to a change in PI. In I16, we split the \xmm~data
into four segments: orbit 1a, orbit 1b, orbit 2a and orbit 2b (see
Fig. 1 in I16). This was to allow for a minor change in instrumental
setup between orbits 2a and 2b, and for possible changes in the source
geometry over the course of orbit 1. We employ the same naming
conventions in this paper. \nustar~observed the source (obs ID
80001044004) simultaneous with the second \xmm~orbit. We use data from
both of the \nustar~focal plane modules, FPMA and FPMB. Here, we use
exactly the same data reduction procedure described in I16.

In this paper, we use only the \nustar~data and the orbit 2
\xmm~data. We keep the data for orbits 2a and 2b separate, but tie
together parameters in our fits between the two \xmm~segments and the
\nustar~exposure. We exclude orbit 1 for a number of reasons. Firstly,
there is no simultaneous \nustar~coverage. It is very useful to
jointly fit \xmm~and \nustar~data, since \xmm~provides high
signal-to-noise at the iron line and \nustar~gives a view of the
reflection hump. Unfortunately, there are some cross-calibration
issues between \xmm~and \nustar, which we will discuss in the
following Section. We show that these issues can be overcome if data
from the two observatories are simultaneous, but without this
simultaneity it is ambiguous whether differences in the spectrum are
down to cross-calibration, or genuinely down to evolution of the
spectrum between the two observations. Secondly, we saw in I16 that
orbits 1a, 2a, 2b and the \nustar~observation all showed the same
characteristic modulation in line energy with QPO phase, with maxima
at $\sim 0.2$ and $\sim 0.7$ cycles, whereas orbit 1b showed a
different modulation. The reason behind this difference is still not
clear, so we exclude the anomalous data set, and also orbit 1a to
avoid simply `cherry picking' the best data.

\subsection{Summary of phase-resolving method}
\label{sec:method}

The phase-resolving method used is described extensively in I16 and
IK15. Here, we summarise the method and leave the details to earlier
references. Conceptually, the method consists of constraining the
average QPO Fourier transform (FT) for each energy channel. That is,
for each energy channel, we wish to measure the mean count rate, and
the amplitude and phase of each observed QPO harmonic. We detect only
the fundamental (first harmonic) and the overtone (second harmonic)
over the broad band noise, and so cannot consider any higher
harmonics. The zeroth harmonic is simply the mean count rate, and also
needs to be taken into consideration. Since the mean count rate is
real, it trivially has a phase of zero. Therefore, for a given energy
channel, the QPO FT consists of five numbers: the (amplitude of the)
mean count rate, the amplitude of the first and second harmonics, and
the phase of the first and second harmonics. Equivalently, we can
think in terms of real and imaginary parts instead of amplitude and
phase, in which case the five numbers are: the real part of the mean
count rate, first and second harmonics and the imaginary part of the
first and second harmonics.

It is fairly straight forward to measure the amplitude of the two QPO
harmonics as a function of energy. The simplest way is to make a power
spectrum for each energy channel and fit each power spectrum with a
sum of Lorentzian functions. The squared
amplitude of the $j$th harmonic is simply the integral of the
Lorentzian component representing it. We used a slightly more
complicated method in I16 to maximize signal-to-noise and to
circumvent the \nustar~deadtime but, conceptually, our method is the
same as described here. The phase can be measured by defining a
reference band, and, for each energy channel, calculating the
cross-spectrum between that channel (the subject channel) and the
reference band (\citealt{vanderKlis1987,Uttley2014}). The phase of the
cross-spectrum averaged over the width of the $j$th harmonic tells us
by how many radians the $j$th harmonic of the subject channel lags the
$j$th harmonic of the reference band. However, we instead want to know
by how many radians the $j$th harmonic of the subject channel lags the
\textit{first} harmonic of the reference band. To know this, we must
measure the phase difference between the first and second harmonics in
the reference band, and use this to correct the phase as measured from
the cross-spectrum. We measure this phase difference between harmonics
using the method of IK15.

With the QPO FT as a function of energy constrained, we can proceed in
two ways. The simplest is to inverse FT the data, to give a waveform
for each energy channel. It is simple to picture this: the waveform
for a given channel is a constant (the mean count rate of that
channel) plus two sine wave functions of QPO phase representing the
two harmonics, each with their own amplitude and phase. Having a
waveform for each energy channel, we can simply take the spectrum for
different values of QPO phase and fit each phase with a spectral
model, and see how the parameters of the spectral model change with
QPO phase. However, the statistics are badly behaved in this
case. This is because, even though the five numbers per energy channel
that make up the QPO FT \textit{do} have well-behaved, Gaussian
errors, the inverse FT introduces correlations between QPO phases. It
is therefore much cleaner from a statistical point of view to define a
model for how the spectrum changes as a function of QPO phase, and
then FT the \textit{model}. Note that both of these methods are
equivalent - we can either inverse FT the data and fit in the time
domain, or we can FT the model and fit in the Fourier domain. The only
difference is that the Fourier domain method is superior when it comes
to assessing goodness of fit, error calculations etc. Therefore, in
this paper, we fit entirely in the Fourier domain.

\section{Time-averaged spectral fits}
\label{sec:cal}

Before analysing the QPO FT, we first fit the time-averaged \xmm~and
\nustar~spectra with a relativistic reflection model. One motivation
for this is to gain insight into the effect of spectral variability on
the time-averaged spectrum. If spectral variability were exclusively
linear, the QPO phase-averaged spectrum would be exactly equal to the
spectrum calculated using the phase-averaged spectral
parameters. However, we saw in I16 that the iron line centroid energy
and the photon index change systematically with QPO phase. These
changes are mildly non-linear and therefore may introduce biases into
time-averaged spectral modelling. Another motivation is to explore the
cross-calibration discrepancy between \xmm~and \nustar. The
\xmm~spectrum is significantly harder than the \nustar~spectrum.

\begin{table*}
	\centering
	\caption{Best fit parameter values and $1\sigma$ errors for
          our time-averaged spectral fit. $\Delta\Gamma$ is a
          calibration parameter that accounts for the offset in
          spectral index measured by \xmm~and \nustar. $R_g$ is a
          gravitational radius, $R_g=GM/c^2$. See the text
          for more details.}
	\label{tab:calfit}
	\begin{tabular}{cccccccccc} 
		\hline
		  Parameter & \vline & $\Delta\Gamma$ & $\Gamma$ &
                $r_{\rm in}$ & $i$ & $E_{\rm cut}$ & $\log_{\rm
                                                     10}\xi$ & $f$ &
                                                                     $A_{\rm Fe}$ \\
		  Units & \vline & & & $R_g$ & deg & keV & & $\%$ & \\
		\hline
		Best fit & \vline & $0.22$ & $1.57$ &
                                                                     $36.3$ & $67.5$ & $285.0$ & $2.0$ & $24.7$ & $0.7$ \\
		$1\sigma$ error & \vline & $5\times 10^{-4}$ & $10^{-3}$ & $4.1$ & $2.3$ & $6.4$ &
                                                                       $5\times 10^{-2}$ & $0.3$
                & $3\times 10^{-2}$ \\
	\end{tabular}
\end{table*}

We consider \xmm~orbit 2a and select a strictly simultaneous interval
of the \nustar~observation. We use \xspec~v12.8.2 to fit the model
\begin{equation}
constant~\times~E^{\Delta\Gamma}~\times~tbabs~\times~[~relxill~+~xillver~],
\label{eqn:specmod}
\end{equation}
to the FPMA, FPMB ($4-75$ keV) and EPIC pn ($4-10$ keV) spectra
simultaneously. We choose to ignore the $< 4$ keV range in the
  \xmm~data in order to avoid calibration features resulting from
  uncertain modelling of the so-called charge transfer inefficiency
  (see \citealt{DeMarco2016} and references therein for a discussion
  on this). The $> 4$ keV energy range also reliably has a negligible
  contribution from direct disc emission, with the disc temperature
  for this observation measured to be $T_{\rm in} < 0.4$ keV by both
  \cite{Stiele2016} and \cite{DeMarco2016}.
The constant factor simply accounts for differences in
absolute flux calibration. We introduce the parameter $\Delta\Gamma$
to account for the discrepancy in photon index measured individually
for the two observatories. We fix $\Delta\Gamma=0$ for both
\nustar~modules, and allow it to go free for the pn. We choose to fit
this way around for a number of reasons. Firstly, the reflection
models we use are only tabulated for $\Gamma \geq 1.4$, so using the
raw \xmm~spectrum, which is very hard, risks going close to this
boundary at some point during the running of the $\chi^2$ minimisation
algorithm. Secondly, the \xmm~spectrum is far harder than is commonly
observed in the hard state (see e.g. \citealt{Shaposhnikov2009}),
whereas the photon index measured from the \nustar~spectrum is
consistent with expectation. Finally, it has previously been reported 
that the pn in timing mode also measures a harder power-law index
than \rxte~for a hard state observation of GX339-4 similar in
flux and spectral shape to the observations analysed here
(\citealt{Kolehmainen2014}). We tie all other parameters to be the
same for the two observatories.

\begin{figure}
	\includegraphics[angle=-90,width=\columnwidth]{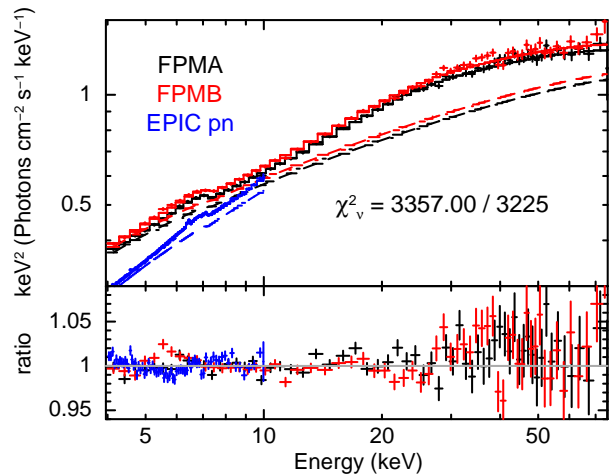}
\vspace{0mm}
 \caption{Unfolded time-averaged spectrum with the best fit model
   (top) and data to model ratio (bottom). We consider
   \xmm~orbit 2a and the strictly simultaneous portion of the
   \nustar~observation. FPMA, FPMB and EPIC-pn data and corresponding
   models are as labelled. The dashed lines represent the continuum
   model. The pn spectrum is significantly harder than the
   \nustar~spectra, but we are able to find an acceptable fit using a
   multiplicative correction factor (equation \ref{eqn:specmod}). Data
   have been re-binned for plotting purposes}
 \label{fig:cal}
\end{figure}

$Tbabs$ accounts for interstellar absorption, and we freeze the
hydrogen column density to $N_h = 2 \times 10^{22} {\rm cm}^{-2}$
assuming the abundances of \cite*{Wilms2000}, following other spectral
analyses of the same data set (\citealt{DeMarco2016,Stiele2016}). This
is slightly higher than the value used in I16, but we find that it has
no significant effect on the measurement of an iron line centroid
energy modulation. $Relxill$ is a relativistic reflection model which
includes an exponentially cut-off power-law X-ray continuum and a
reflection component which is smeared by the orbital motion of disc
material and gravitational redshift (\citealt{Garcia2014}). We include
the $xillver$ component (\citealt{Garcia2013}), which is the same as $relxill$ but without
the relativistic smearing, to account for a distant reflector which it
is possible to detect in \xmm~spectra of BH X-ray binaries in
the form of a narrow iron line (e.g. Cygnus X-1: \citealt{Fabian2012};
GX 339-4: \citealt{Kolehmainen2014}). The shape of the
rest-frame reflection spectrum depends on the shape of the
illuminating continuum, the disc ionisation parameter $\log_{10}\xi$
and the iron abundance relative to solar, $A_{Fe}$. For the $relxill$
component, we allow $\log_{10}\xi$ and $A_{Fe}$ to be free
parameters. For the distant reflector, we assume neutral material
($\log_{10}\xi=0$), with the same iron abundance as the
$relxill$ component. The relativistic smearing depends  on the
inclination angle $i$, the disc inner radius $r_{\rm in}$ and the
radial dependence of illuminating flux, which we assume to be $\propto
r^{-3}$. Since we do not assume that $r_{\rm in}$ is equal to the
innermost stable circular orbit (ISCO), the relativistic smearing only
depends very weakly on the BH spin parameter, $a$ (the energy shifts
and photon paths both depend on the metric). For this reason, we fix
$a=0.21$ to be consistent with the measurement of $a\approx 0.2$ made
by \cite{Steiner2012} through disc spectral fitting and the limit $a
\gtrsim 0.2$ placed by \cite{Ingram2014} using high frequency QPOs.

Fig \ref{fig:cal} shows the data unfolded around the best-fit
model (top) and the ratio of data to model (bottom). After applying $0.5\%$
systematic errors to account for uncertainties in the
  telescope response matrices (as is widely practiced:
  e.g. \citealt{Kolehmainen2014,Plant2015}), we achieve an acceptable
fit (reduced $\chi^2=3357/3225=1.04$). We find through an
  F-test that the distant reflector is formally required with a
  significance of $5.3 \sigma$ (since the best-fit without the
  $xillver$ component has a reduced $\chi^2 = 3387/3226$). Table
\ref{tab:calfit} shows the resulting best-fit parameters with $1
\sigma$ errors. Our best fit model indicates that the disc is
truncated outside of the ISCO and yields a moderately high
inclination, consistent with the source showing dips but not eclipses
(\citealt{Homan2005b}). \cite{Steiner2012} measured the angle between 
our line of sight and the radio jet (which is likely aligned with the
BH spin axis; \citealt{Blandford1977}) in H 1743--322 to be $\sim
75^\circ$. Thus, although the two angles are broadly consistent within
errors, there is room for a modest misalignment between the disc and
BH spin axes, as is required for the precession model. The fit
indicates a fairly low ionisation, consistent with the relatively low
continuum luminosity, and a mildly sub-solar iron abundance.

\section{Tomographic model}
\label{sec:model}

\begin{figure*}
	\includegraphics[clip=true, width=18 cm,trim= 0.0cm
        6.2cm 1.0cm 5.0cm]{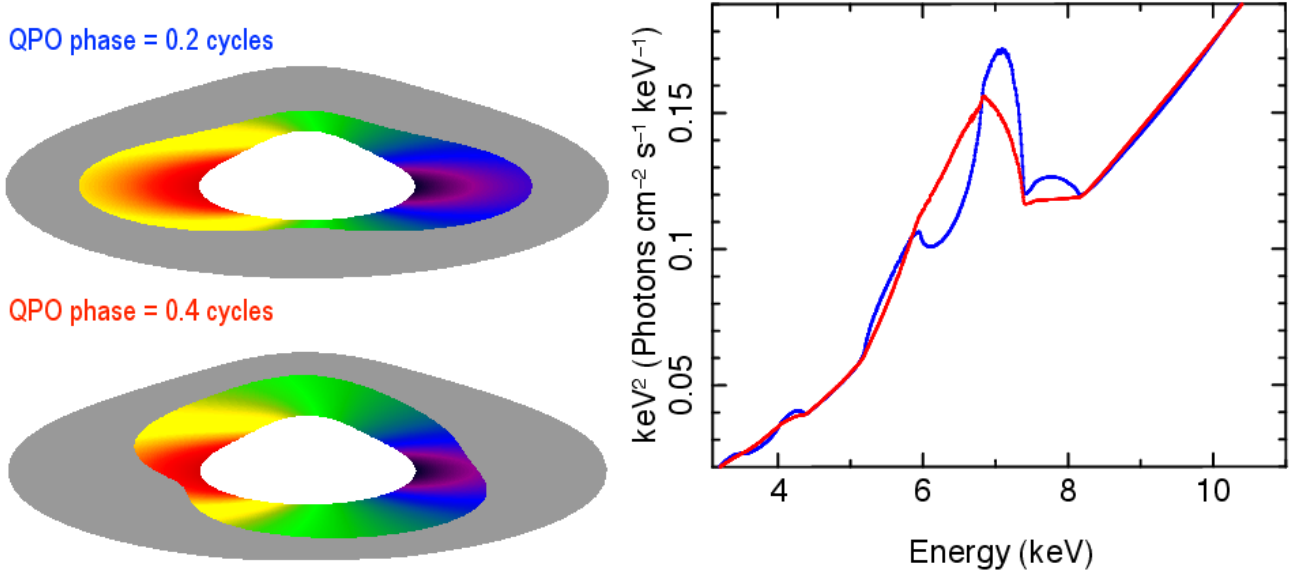}
\vspace{-7mm}
 \caption{\textit{Left:} Ray traced disc images for two QPO phases (as
   labelled) visualising our best fit model. The disc inner radius and
   inclination are equal to our best fit values, and the border of
   the multi-coloured patches is set by the best fit illumination profile for
   the relevant QPO phase. The colour scheme of the multi coloured
   patches encodes blue shifts. \textit{Right:} The best fitting reflection spectrum,
   zoomed in on the iron line, for the same QPO phases. The QPO phase
   = $0.2$ cycles iron line (blue) has a boosted blue horn and red
   wing because the blue and red shifted parts of the disc are
   preferentially illuminated (top image), in contrast to the QPO
   phase = $0.4$ iron line (red line), which corresponds to the bottom
   disc image. Animated versions of these plots can be viewed at and
   downloaded from \color{blue}\underline{\smash{https://figshare.com/articles/Tomographic\_modelling\_of\_H\_1743-322/3503933}}\color{black}. These
  animations are designed to be played together.}
 \label{fig:profiles}
\end{figure*}

In this section, we describe our physical model for the QPO
phase-resolved reflection spectrum. The intention is to mimic the
asymmetric, rotating illumination pattern that would be caused by a
precessing inner flow irradiating the disc by using an analytic
parameterisation which we can fit directly to the data. The bright
patches on the disc in Fig. \ref{fig:profiles} illustrate this for two
QPO phases. We represent the reflected intensity as a function of disc
radius $r$, disc azimuth $\phi$, and QPO phase $\gamma$ using the
function
\begin{eqnarray}
I_{E_e}(r,\phi,\gamma) \propto &r^{-q}& \Big\{ 1 + A_1 \cos^2\left[ (
\gamma-\phi+\phi_1 ) / 2
\right] 
\nonumber \\
&+& A_2 \cos^2\left[ \gamma-\phi+\phi_2 \right]
\Big\}  I_{E_e},
\label{eqn:illum}
\end{eqnarray}
where $I_{E_e}$ is the rest-frame reflection spectrum and
$I_{E_e}(r,\phi,\gamma)$ is the specific intensity of radiation
emitted at photon energy $E_e$ for a given patch of the disc at a given
QPO phase. We see that the intensity is a power-law function of
radius, the same as our fits to the time-averaged spectrum in Section
\ref{sec:cal} (with $q=3$). We define the z-axis to be parallel with
the disc rotation axis such that the disc lies in the x-y plane. The
disc azimuth $\phi$ is measured clockwise from the x-axis, which is
defined as the projection of the observer's line-of-sight on the black
hole equatorial plane. The dependence of intensity on disc azimuth is
parameterised through the cosine terms in equation
\ref{eqn:illum}. Setting $A_2=0$, $A_1>0$ creates only one bright
patch that rotates about the disc surface once per precession cycle,
leading to only one maximum in the iron line energy per QPO cycle. At
QPO phase $\gamma=0$, the brightest patch of the disc in this case
would be $\phi=\phi_1$. Setting $A_1=0$, $A_2>0$ corresponds to the
front and back of the flow irradiating the disc with equal intensity,
leading to two identical patches rotating about the disc surface and
two identical maxima in the line energy per QPO cycle. At QPO
phase $\gamma=0$, the two brightest patches of the disc in this case
would be $\phi=\phi_2$ and $\phi=\phi_2+180^\circ$. In I16, we
observed two maxima in the iron line centroid energy per QPO cycle
with the second slightly higher than the first, thus the best fit
model will likely have $A_1>0$, $A_2>0$. This indicates that the front
and back of the flow irradiate the disc, and thus requires the flow to
have a fairly small vertical extent (or alternatively the misalignment
between disc and flow is large). \cite{Ingram2012a} instead
modelled the flow as an oblate spheroid with large vertical extent,
and therefore the underside of the flow was never above the disc
mid-plane. The cosines in equation \ref{eqn:illum} are squared so as
to prevent the possibility of unphysical parameter combinations for
which the intensity is negative.

The specific flux observed at energy $E_o$ from a patch of the disc
subtending solid angle $d\Omega(r,\phi)$ to the observer is
\begin{equation}
dF_{E_o}(r,\phi,\gamma) = (E_o/E_e)^3 I_{E_e}(r,\phi,\gamma)  d\Omega(r,\phi).
\label{eqn:dF}
\end{equation}
Each solid angle element $d\Omega(r,\phi)$ can be seen as a pixel on
the observer's camera. We define a $400 \times 400$ circular
(polar) grid of pixels and, for each one, trace the unique
null-geodesic in the Kerr metric (using the code \textsc{geokerr};
\citealt{Dexter2009}) back from the centre of the pixel to assess if
and where that geodesic intercepts the disc. The blue shift,
$(E_o/E_e)$, is calculated for all the photon paths that intercept the
disc using the equation
\begin{equation}
\frac{E_o}{E_e} = \frac{ \sqrt{ -g_{tt} - 2g_{t\phi} \omega -
    g_{\phi\phi} \omega^2  } } {1 + \omega \alpha \sin(i)},
\label{eqn:gfac}
\end{equation}
where $g_{\mu \nu}$ is the Kerr metric, $\omega=1/(r^{3/2}+a)$ is the
angular velocity in dimensionless units, and the impact parameter
$\alpha$ is the horizontal distance from the centre of the observer's
camera to the centre of the pixel in $R_g$. We use the same coordinate
system and ray-tracing procedure as \cite{Middleton2015} except here
we assume clockwise (left-handed) rotation so that blue shifts are
always to the right and red shifts are always to the left both for
disc images and spectra (see Fig. \ref{fig:profiles}). \footnote{The original derivation  of equation
  \ref{eqn:gfac}, presented in \cite{Luminet1979} (equation 18
  therein), contains a typographical error, which was repeated in
  equation A3 of \cite{Middleton2015} but corrected here (the results
  of \citealt{Middleton2015} were calculated using the correct
  formula).} Equation \ref{eqn:gfac} assumes disc rotation in the BH equatorial plane. This is not true if there is a
misalignment between disc and BH rotational axes, as we are
imagining here, but the error introduced is very small
(\citealt{Ingram2015a}). We use $xillver$ (\citealt{Garcia2013}) for
the rest-frame reflection spectrum, $I_{E_e}$ (see section
\ref{sec:cal}).

We allow the continuum normalisation $N$, the reflection fraction $f$, and
the photon index, $\Gamma$ to vary as a function of QPO phase,
$\gamma$, with two non-zero harmonics. For example, the photon index
varies with QPO phase as
\begin{equation}
\Gamma(\gamma) = \Gamma_0 + A_{1\Gamma} \sin[ \gamma - \phi_{1\Gamma} ] + A_{2\Gamma}
\sin[ 2(\gamma - \phi_{2\Gamma}) ],
\label{eqn:Gamma}
\end{equation}
and we use analogous expressions for the other two variable parameters,
$N(\gamma)$ and $f(\gamma)$.

Our model calculates the integrated specific flux as a function of
observed energy $F_{E_o}(\gamma)$ for 16 steps in $\gamma$ and, for
each energy channel, calculates the FT $\tilde{F}_{E_o}(j)$
for the harmonics $j=0$, $j=1$ and $j=2$. The $j=0$ harmonic (the
so-called DC component) is simply the spectrum averaged over all
$\gamma$. Each harmonic has real and imaginary parts, but the
imaginary part of the DC component is trivially zero. This leaves five
spectra to fit simultaneously to the observed QPO FT. We load our
model into \textsc{xspec} using the local model 
functionality. We include absorption using $tbabs$ for real and
imaginary parts of every harmonic (since absorption is
multiplicative). We also add a distant reflector ($xillver$) to the DC
component only, since a constant additive component does not
contribute to the non-zero harmonics. As discussed in I16, the fact
that we fit in terms of real and imaginary parts rather than amplitude
and phase means that the instrument response is trivially accounted
for when the data are loaded into \xspec. In order to avoid edge
effects of the convolution, we extend the energy grid used to
calculate the model up to $200$ keV (using the \xspec~command
`$energies~extend~high~200$').

\begin{figure*}
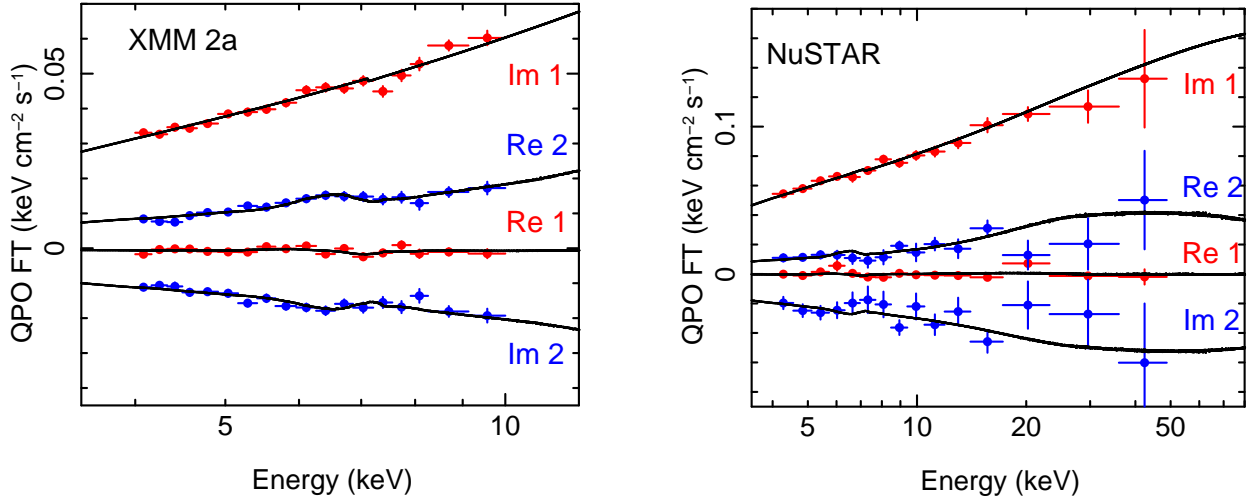

	\includegraphics[angle=0,width=\columnwidth]{xmmQPOFT.ps} ~~~~~
	\includegraphics[angle=0,width=\columnwidth]{nustarQPOFT.ps}
\vspace{-7mm}
 \caption{QPO FT as a function of energy for \xmm~orbit 2a (left) and
\nustar~(right). Real and imaginary parts of the first and second
harmonics are as labelled. Real and imaginary parts of the data are
colour coded blue and red respectively, and the black lines depict the
best fitting model. The data are unfolded around the best fitting
model and are in units of energy squared $\times$ specific photon
flux. We see features around the iron line and reflection hump,
particularly for the real parts. \xmm and \nustar show the same
trends.}
 \label{fig:qpoft}
\end{figure*}

\section{Results}
\label{sec:results}

We load the model described in the previous section into \xspec~as the
local model $modfeprec$. We fit to the measured energy dependent
FT of the QPO, considering the real and imaginary parts
of the first and second harmonics, and also the time-averaged spectrum
(which is the real part of the zeroth harmonic). Altogether, this
gives five spectra to simultaneously fit for each data set. For the non-zero
harmonics, we use the same coarse binning employed in I16, which
is necessary to ensure Gaussian errors. For the zeroth harmonic (the
time-averaged spectrum), we instead use the fine spectral binning
employed here in Section~\ref{sec:cal} (although note that this is
still coarser than the instrument response). As a check, we also performed
fits using coarse binning for the time-averaged spectrum and see no
significant differences in our best-fit parameters or goodness of
fit. We use the fine binning for our analysis, however, since it is
sensitive to parameter combinations that can reproduce the observed
variability properties but predict sharp features in the time-averaged
spectrum that are not observed.

We employ the model
\begin{equation}
constant~\times~E^{\Delta\Gamma}~\times~tbabs~\times~[~modfeprec~+~xillver~],
\end{equation}
where $modfeprec$ outputs either the real or imaginary part of the
first, second or zeroth QPO harmonic. We fix the calibration parameters
($constant$ and $\Delta \Gamma$) to the values obtained in Section
\ref{sec:cal}, and continue to use $N_h=2\times 10^{22}$ cm$^{-2}$ for
the hydrogen column density. We also fix the ionisation parameter,
high energy cut-off and relative iron abundance to the values obtained
in Section \ref{sec:cal}. The $xillver$ component, as before, accounts
for distant reflection, which we assume not to vary on the QPO period
and therefore set its normalisation to zero for all non-zero
harmonics. We jointly fit for the two \xmm~observations and the
\nustar~observation. We tie all physical parameters between these
three data sets, except for those describing the modulation of the
continuum normalisation. As in I16, the modulation in the continuum
normalisation is very similar for the three data sets, but can be
measured to a high enough precision for small differences to be highly
significant.

\subsection{Best fit tomographic model}

We achieve a good fit with reduced $\chi^2=2544.54/2511=1.013$
(rejection probability $68.5\%$). Fig. \ref{fig:qpoft} shows the
QPO FT data and model for \xmm~orbit 2a (left) and
\nustar~(right). Here, the real and imaginary parts of the first and
second harmonics are as labelled, with real and imaginary data points
colour coded respectively blue and red, and the model always plotted
as a black line. The data are unfolded around the instrument response
assuming the best fit model and are in units of energy squared
$\times$ specific photon flux (i.e. the $eeuf$ option in \xspec). We
see curvature around the iron line, particularly for the real
parts. In the \nustar~data, we also see the effect of the reflection
hump at high energy. In the model, these features result from changes
in the shape of the reflection spectrum over a QPO cycle.

\begin{table*}
	\centering
	\caption{Best fit parameters for our tomographic modelling. See the text for
more details.}
	\label{tab:bestfit}
	\begin{tabular}{cccccccccc} 
		\hline
		  Parameter & \vline & $\Gamma_0$ &
                $r_{\rm in}$ & $i$ & $f_0$ & $A_1$ & $A_2$ & $\phi_1$ & $\phi_2$ \\
		  Units & \vline & & $R_g$ & deg & $\%$ & &
                  & deg & deg \\
		\hline
		$1\sigma$ upper & \vline & $1.571$ & $37.30$ & $73.33$ & $23.20$ & $2.263$ & $8.740$ & $138.685$ & $28.105$ \\
		Best fit & \vline & $1.567$ &
 $31.47$ & $70.68$ & $21.83$ & $0.93$ & $3.50$ & $94.66$ & $18.98$ \\

		$1\sigma$ lower & \vline & $1.563$ & $27.81$ & $67.51$ & $20.48$ & $0.322$ & $1.275$& $49.675$ & $8.93$ \\
		\hline
	\end{tabular}
\end{table*}

Table \ref{tab:bestfit} shows our
best-fit parameters with $1~\sigma$ errors. Fig. \ref{fig:profiles}
shows a visualisation of the disc illumination profile indicated by
the best-fit asymmetry parameters $A_1$, $A_2$, $\phi_1$ and $\phi_2$
(see equation \ref{eqn:illum}). We show disc images for two QPO phases
with the corresponding phase-resolved reflection spectra. The
multi-coloured patches pick out where the illuminating intensity,
$I_{E_e}(r,\phi,\gamma)$, is greater than $10\%$ of its maximum value,
with the rest of the disc coloured grey. The colour coding of the
patches encodes blue shifts (equation \ref{eqn:gfac}). The two QPO phases
shown are $0.2$ cycles ($72^{\circ}$, blue line profile) and $0.4$
cycles ($144^{\circ}$, red line profile), since these roughly
correspond respectively to the maximum and minimum line centroid
energy, as measured by I16. For the purposes of these plots, the
modulations in $\Gamma$, reflection fraction and normalisation have
been set to zero, to ensure that all changes to the reflection
spectrum result purely from changes to the disc illumination
profile. All other parameters come directly from our best fitting
model. We see that the line has a strongly boosted blue horn and a
suppressed core (blue line) when the left and right sides of the disc
are illuminated (top disc image). This is because we see enhanced
emission from both the blue shifted approaching material and the red
shifted receding material, but Doppler boosting ensures that the blue
shifted emission dominates. In contrast, the line has a strong core
but suppressed wings (red line) when the front and back of the disc
are illuminated (bottom disc image), since there is no enhancement of
the Doppler shifted emission from the approaching and receding disc
material. As with the time-averaged fits, we measure a
moderately truncated disc and a fairly high inclination angle. Light
bending effects are evident in the disc images through apparent
warping of the disc, but we do not consider ghost images, since
photons on these paths will likely be scattered before reaching the
observer. Animated versions of the plots shown in Fig
\ref{fig:profiles} can be viewed at and downloaded from the link given
in the caption.

\begin{figure}
\center
	\includegraphics[angle=0,width=\columnwidth]{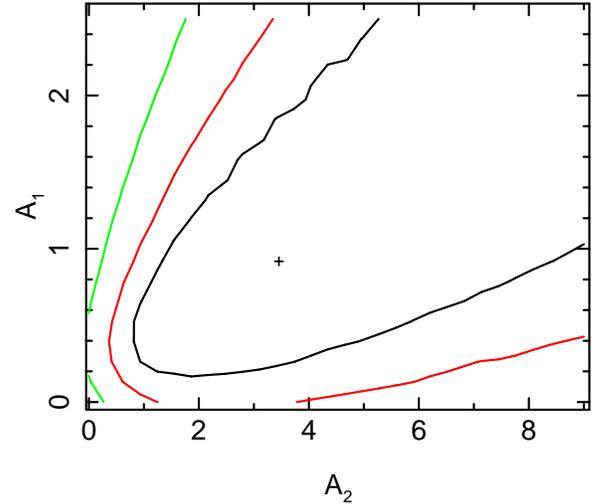}
\vspace{-10mm}
 \caption{$\chi^2$ contour plot showing the two asymmetry parameters $A_1$ and
$A_2$. The contours correspond to 1 (black), 2 (red) and 3 (green)
$\sigma$ confidence for two degrees of freedom.}
 \label{fig:contours}
\end{figure}

Fig. \ref{fig:contours} is a contour plot resulting from varying the
asymmetry parameters $A_1$ and $A_2$. The contours represent $\Delta
\chi^2=2.3$ (black), 6.18 (red) and 11.83 (green). These levels
correspond to 1, 2 and 3 $\sigma$ for two degrees of freedom. We see
that fairly large values of $A_1$ and $A_2$ return a $\chi^2$ value
within 1$\sigma$ of the best-fit (black cross). This is because of the
way the illuminating flux, $I_{E_e}(r,\phi,\gamma)$, is
parameterised. We see from equation \ref{eqn:illum} that
$I_{E_e}(r,\phi,\gamma)$ is a sum with three terms. The first term
does not depend on QPO phase, whereas the second two do. Increasing
$A_1$ and $A_2$ respectively increases the relative importance of the
second and third terms with respect to the first. However, increasing
$A_2$ from, e.g. $100$ to $1000$ makes a negligible difference, since the
constant component changes from being $1\%$ to $0.1\%$ of the flux in
this case. Our parameterisation is of course designed to investigate
the effect of setting $A_1=A_2=0$, in which case there is no QPO phase
dependence of the line profile in the model. We see that $A_2$ is
better constrained than $A_1$ and that the point $A_1=A_2=0$ lies
outside of the $3\sigma$ contour. However, these contours are for two
degrees of freedom, and the point $A_1=A_2=0$ is a special point, in
that setting $A_1=A_2=0$ renders the fit insensitive to $\phi_1$ and
$\phi_2$. Following I16, we therefore use an F-test to compare our best
fit (reduced $\chi^2=2544.54/2511$) with the null-hypothesis (reduced
$\chi^2=2556.98/2515$). This indicates that the best-fit model is
preferred with $2.40~\sigma$ confidence. Therefore, although we can
say with $3.70\sigma$ confidence that the line centroid energy is
modulated (I16), our analysis yields a lower significance for
an actual asymmetric illumination profile. This is partly because here
we necessarily use a more complex model for the iron line than a
Gaussian, and therefore lose degrees of freedom. Also, small apparent
shifts in the iron line profile can be driven by changes in the
reflection continuum, caused by changes in the photon index (which our
model automatically takes into account). Finally, we have been rather
conservative in excluding \xmm~orbit 1.

\begin{figure}
	\includegraphics[angle=0,width=\columnwidth]{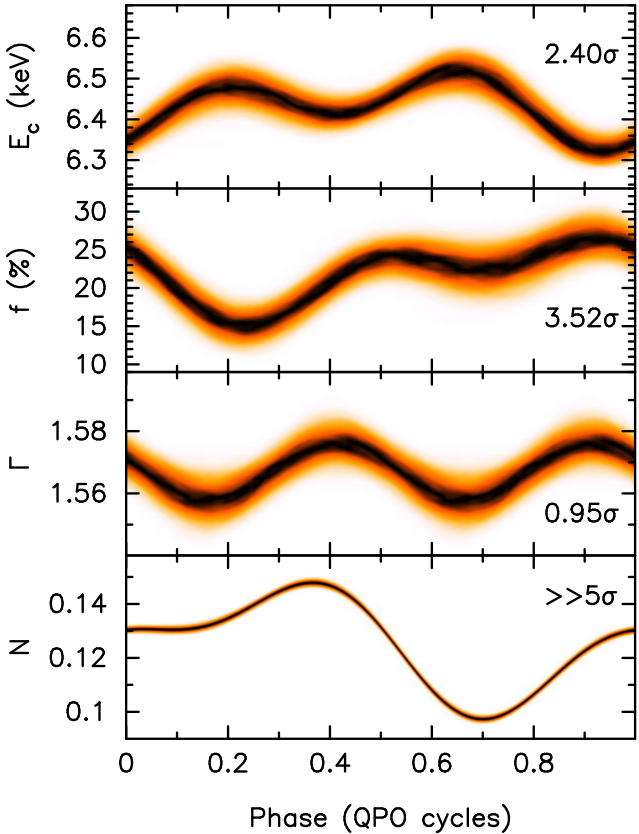}
\vspace{-5mm}
 \caption{Visualisation of our best fitting parameter modulations. The
   iron line centroid energy, defined by equation \ref{eqn:centroid},
   shows a characteristic variation with QPO phase, with two maxima at
   $\sim 0.2$ and $\sim 0.4$ cycles. This is the same trend as is
   presented in I16 for a Gaussian iron line model. We also see that
   the reflection fraction is modulated ($3.52\sigma$) and the photon
   index modulation is not statistically significant. The
   modulation in normalisation corresponds to \xmm~orbit 2a.}
 \label{fig:paras}
\end{figure}

As another visualisation of our results,
Fig. \ref{fig:paras} shows how various quantities / model parameters
vary with QPO phase. Each panel is labelled with a statistical
significance. For the top panel, this is the significance of
asymmetric illumination as calculated above. For the other three
panels, this is the significance with which the plotted parameter
changes with QPO phase, calculated using an F-test comparing the best
fitting model with an alternative fit whereby the parameter in
question is forced to be constant. In order to make this plot,
we run a Monte Carlo Markov Chain for our best fit model and,
for each of the four panels, calculate a histogram for each of 512
QPO phases (see I16 for details). The chain has 125,000 steps and we
burn the first 25,000. For the top panel, we compute the line profile
as a function of QPO phase, $L(\gamma,E)$, assuming a
$\delta$-function rest-frame iron fluorescence line at $6.4$ keV and
define the centroid energy as
\begin{equation}
E_{\rm c}(\gamma) = \frac{ \int_0^\infty E L(\gamma,E) {\rm d}E }{ \int_0^\infty L(\gamma,E) {\rm d}E }.
\label{eqn:centroid}
\end{equation}
We see that the centroid energy calculated in this way follows the
same trend as the centroid of the Gaussian used in I16, with maxima at $\sim
0.2$ and $\sim 0.7$ cycles. The second panel shows the reflection
fraction. We see that this is modulated with QPO phase
($3.52\sigma$). The significance is calculated from an F-test, and
does not depend on the chain. Our model is normalised such that the observed
reflected flux will be higher when the approaching disc material is
preferentially illuminated, even if $f$ remains constant, because of
Doppler boosting. Changes in $f$ therefore indicate variability in the
reflected flux \textit{on top} of this effect. We expect changes in
the reflection fraction if the misalignment angle between the inner
flow and disc changes throughout a precession cycle
(\citealt{Ingram2012a}). The third panel shows a modulation in
$\Gamma$, however this is not significant ($0.95\sigma$). This
modulation in $\Gamma$ has a different phase to that measured by I16,
with maxima at $\sim 0.4$ and $\sim 0.9$ cycles, compared with $\sim
0.3$ and $\sim 0.8$ cycles in I16. The significance of the $\Gamma$
modulation has also drastically reduced compared with I16. This is
because we now include a full reflection model with a continuum rather
than just a Gaussian iron line. Our results indicate that the observed
changes in spectral hardness during a QPO cycle are more due to
changes in reflection fraction than photon index. An increase in
reflection fraction makes the spectrum harder (since the reflected
continuum is harder than the directly observed continuum). With no
reflected continuum in the model, this hardening can only be modelled
as a reduction in $\Gamma$. We can see evidence of this in
Fig. \ref{fig:paras}, since the lowest reflection fraction ($\sim 0.25$
cycles) coincides with the highest $\Gamma$ in I16. The \nustar~data
is particularly important for constraining this, since the reflection
hump gives a good constraint on the reflected continuum.

%
For completeness, we briefly investigate the anomalous \xmm~orbit
1b. In I16, we found that the best-fit iron line centroid energy
modulation was very different in this data set to all the others, and
also that the modulation was not statistically significant (see Fig. 7 in
I16). As expected, when we fit these data with our tomographic model, the
asymmetry parameters $A_1$ and $A_2$ are poorly constrained, and the
best-fit model with $A_1=0.96$ and $A_2 \approx 0$ is preferred to
$A_1=A_2=0$ with a significance of only $0.5 \sigma$. The absence of
simultaneous \nustar~data for this data set adds to the difficulty in
constraining model parameters. Our best fitting value of $A_2 \approx 0$
is consistent with the results of I16, who found that the best fitting
iron line centroid energy modulation had no second harmonic
(i.e. $A_{2E} \approx 0$). Our best fitting value of $\phi_1 = 160.9^\circ$
is also consistent with the I16 results, since this implies that the
line centroid energy peaks at $\sim 0.5$ QPO cycles (see Fig. 7 in
I16) - very different to the other data sets.

\subsection{Alternative models}
\label{sec:alt}

We also consider alternative interpretations for the iron line
centroid energy modulation.

\subsubsection{Modulated ionisation parameter}
We first consider changes in the ionisation parameter, $\log_{10}\xi$,
over a QPO cycle. An increase in ionisation leads to a higher
rest-frame line energy, since the ions are on average more tightly
bound (\citealt{Matt1993}; \citealt{Done2010a}). Therefore, it is
possible, in principle, for there to be a modulation in line energy
with no geometric changes. In this case, the disc ionisation should
peak when the irradiating flux peaks, since it is the intensity of
incoming radiation that governs the ionisation balance. Thus, we would
expect in this case that the line energy should be in phase with the
continuum flux, which is not the case
(Fig. \ref{fig:paras}). Nonetheless, we test the ionisation hypothesis
without this constraint. We set $A_1=A_2=0$ and parameterise the
ionisation parameter as a function of QPO phase,
$\log_{10}\xi(\gamma)$, in the same way as $\Gamma(\gamma)$ in
equation \ref{eqn:Gamma}, with the average, amplitudes and phases
replaced by $\log_{10}\xi_0$, $A_{1\xi}$, $A_{2\xi}$, $\phi_{1\xi}$
and $\phi_{2\xi}$.

The null-hypothesis, with $A_{1\xi}=A_{2\xi}=0$ has
a reduced $\chi^2=2556.98 / 2515$, and the best fit we find after
releasing $A_{1\xi}$ and $A_{2\xi}$ has reduced $\chi^2=2556.96 /
2511$. The negligible change in $\chi^2$ for a reduction of 4 degrees
of freedom means that this model is not an improvement over the
null-hypothesis. Our best fit tomographic model is preferred over this
alternative model, but the significance of this cannot be measured
using an F-test since the two models have the same number of degrees
of freedom. We calculate a lower limit of the significance through an
F-test by artificially adding a degree of freedom onto the alternative
model. From this, we conclude that the our best fit tomographic model
is preferred to the alternative model with a significance of $>
3.5\sigma$ (see Table \ref{tab:comp}). This model does not work
because increasing ionisation increases the line energy and
\textit{suppresses} the relative strength of the reflection hump,
which is the complete opposite of what we observe (see Fig. 3 in
I16).

\subsubsection{Modulated disc inner radius}
Our results strongly favour a systematic geometric change
during the QPO cycle. Perhaps an axisymmetric change is adequate
though? We consider a modulation of the disc inner radius with
parameters $r_{{\rm in}0}$, $A_{1r}$, $A_{2r}$, $\phi_{1r}$ and
$\phi_{2r}$. This can cause changes in the line profile because
rotational velocity depends on radius. We again set $A_1=A_2=0$ and
start with the null-hypothesis model $A_{1r}=A_{2r}=0$. When we
release $A_{1r}$ and $A_{2r}$, we find a best fit with reduced
$\chi^2=2556.2 / 2511$. Using the same method as before, we find that
our best-fit model is preferred over this alternative model with $>3.4
\sigma$ significance.

\subsubsection{Modulated emissivity profile}
Finally, we consider a modulation in the radial emissivity
profile. This will also influence the line profile because of the
radial dependence of rotational velocity (and gravitational
redshift). We see in equation \ref{eqn:illum}, that the illuminating
flux is $\propto r^{-q}$. Here, we parameterise $q$ with the
parameters $q_0$, $A_{1q}$, $A_{2q}$, $\phi_{1q}$ and $\phi_{2q}$. The
best fit we find has reduced $\chi^2=2556.9 / 2511$, and so our
best-fit model is preferred over this with $>3.5\sigma$ significance
(see Table \ref{tab:comp} for a comparison of all the models
tested).

\begin{table}
	\centering
	\caption{Comparison of models. The best fitting model is our
          tomographic model and the null-hypothesis model considers
          axi-symmetric illumination of the disc
          (i.e. $A_1=A_2=0$). The three alternative models tested in
          Section \ref{sec:alt} are also listed. In the third column,
          we list the statistical significance with which the best fitting model is
          preferred over each alternative model.}
	\label{tab:comp}
  \bgroup
  \def\arraystretch{1.5}
	\begin{tabular}{lccc} 
		\hline
		  Model & \vline & $\chi^2/$d.o.f. & Significance \\
		\hline
		  Best fit & \vline & $2544.54/2511$ & - \\
		\hline
		  Null-hypothesis & \vline & $2556.98/2515$ & $2.4 \sigma$ \\
		\hline
		 $\xi$ modulation & \vline & $2556.96/2511$ & $> 3.5 \sigma$\\
		\hline
		  $r_{\rm in}$ modulation & \vline & $2556.20/2511$ &$> 3.4\sigma$ \\
		\hline
		  $q$ modulation & \vline & $2556.90/2511$ & $> 3.5 \sigma$\\
		\hline
	\end{tabular}
  \egroup
\end{table}

\section{Discussion}
\label{sec:discussion}

We have developed a spectral model that calculates the reflection
spectrum emitted from a disc with an asymmetric, rotating illumination
pattern. This is designed to mimic the effect of a precessing inner
flow preferentially illuminating different disc azimuths during a
precession cycle, but makes no \textit{a priori} assumptions about the
inner flow geometry. The asymmetry in the illumination profile, and
therefore the QPO phase dependence of the iron line profile, is
parameterised by the asymmetry parameters $A_1$ and $A_2$. We fit this
model, in Fourier space, to the QPO phase-resolved spectra from H
1743--322, originally constrained by I16. In this Section, we discuss
our results.

\subsection{Asymmetric illumination profile}

For our best fit model, $A_1\approx 0.9$ and $A_2\approx 3.5$,
indicating an asymmetric illumination profile that rotates about the
disc surface throughout a QPO cycle. This is visualised in
Fig. \ref{fig:profiles} by the multi-coloured patches. Since $A_2 >
A_1$, there are two bright patches rotating about the disc
surface. The iron line has its maximum centroid energy when the left
and right hand sides of the disc are illuminated (QPO phase $\sim 0.2$
cycles), and it has its minimum centroid energy when the front and back
of the disc are illuminated (QPO phase $\sim 0.2$ cycles). These
configurations both occur twice per precession cycle,
explaining why we see two maxima in line centroid energy per QPO
cycle (top panel of Fig. \ref{fig:paras}; also see Fig. 10 of I16). In
I16, we suggested that such an illumination profile could result from
the disc being irradiated by both the front and back of the precessing
flow. This could occur if the vertical extent of the flow is
relatively small compared with the misalignment between the disc and
flow, since in this case the underside of the flow can be above the
disc. The true configuration is likely more complex than this, perhaps
with a transition region, or even differential precession warping the
inner flow as suggested by \cite*{vandenEijnden2016}.

We find that our best fit model is preferred to a null-hypothesis with
$A_1=A_2=0$ with $2.4\sigma$ confidence. This is a lower significance
than for the iron line centroid energy modulation found by I16
($3.7\sigma$), because we are now fitting a more complex
model with less degrees of freedom, and we also conservatively ignore
$\sim 130$ ks of data. We also fit alternative models for the line centroid energy
modulation. We model modulations in the disc ionisation parameter,
inner radius and radial emissivity profile. We find that none can
explain the observed QPO phase dependence of the iron line. We note
that a model whereby the disc inclination angle changes will likely
provide an acceptable fit. Alternative models considering precession
of the \textit{reflector} rather than the illuminator
(\citealt*{Schnittman2006}) therefore cannot be ruled out.

\subsection{Light-crossing lags}

Our analysis has not considered light-crossing time lags, since they
are small compared with the timescales we are considering. Since we
have simply parameterised the illumination profile on the disc surface
as a function of QPO phase, $I(r,\phi,\gamma)$, light-crossing
lags can, in principle, be swallowed up into our definition of
$I(r,\phi,\gamma)$. We can estimate the importance of light-crossing
lags by imagining that equation \ref{eqn:illum} represents the
illumination under the assumption that light travel is
\textit{instantaneous}. In this case, a patch of the disc located at
$r,\phi$ sees the illumination pattern corresponding to the QPO phase
$\gamma'=\gamma-l(r,\phi)\nu_{qpo}/c$, where $l(r,\phi)$ is the path
length from the illuminating source to the disc patch. Using the
approximation $l \approx r R_g$, the expression for the
illuminating flux becomes $I(r,\phi,\gamma-r \nu_{qpo} R_g/c)$. For
the observation considered here, $\nu_{qpo}\approx 0.25$ Hz, so even
at $r=100$ and assuming $M=10 M_\odot$, this correction to the phase
is only $r \nu_{qpo} R_g/c \sim 10^{-3}$ cycles. If we were
considering instead a QPO with $\nu_{qpo}=25$ Hz, however, we see that
this correction becomes significant at $\sim 0.1$ cycles. Therefore,
analysis of higher frequency Type C QPOs should take
light-crossing lags into account when interpreting the measured QPO phase
dependent illumination profile.

\subsection{Misalignment}

\begin{figure}
	\includegraphics[angle=0,width=\columnwidth]{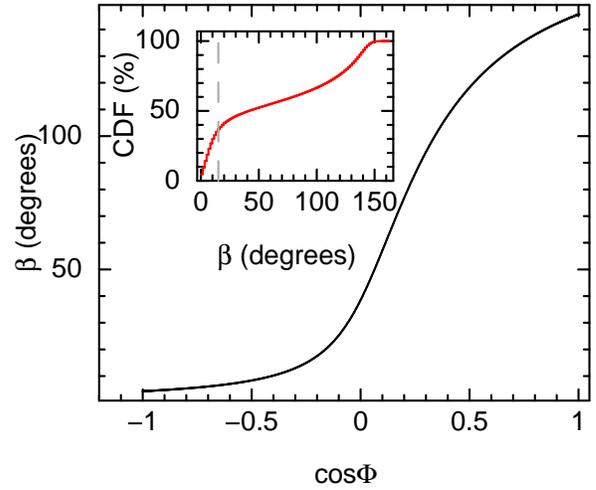}
\vspace{-9mm}
 \caption{\textit{Main plot:} Misalignment angle between the disc and
   BH spin axes, $\beta$, plotted against the cosine of the azimuthal
   viewing angle (see the text for details). This assumes our best
   fitting value for the disc inclination angle $i$ and a best fitting
   value of the BH spin inclination angle $\theta$ from the
   literature. \textit{Inset:} Cumulative distribution function for
   $\beta$ which takes statistical measurement errors of $i$ and
   $\theta$ into account and assumes equal probability of measuring a
   given $\cos\Phi$.} 
 \label{fig:beta}
\end{figure}

In the precession model, the inner flow spin axis is assumed to precess
around the BH spin axis, such that the angle between the BH and flow
spin axes stays constant. Since Lense-Thirring precession does not
occur in the BH equatorial plane, a misalignment between the disc and
the BH spin axes is assumed. This way, the inner flow is being fed by
a misaligned disc, driving precession. Defining the angle between the
disc and BH axes as $\beta$, the angle between the BH and flow axes is
also $\beta$ and the angle between the inner flow and the disc varies
over a precession cycle from a minimum of $0$ to a maximum of $2\beta$
(see schematics in \citealt*{Veledina2013} and
\citealt{Ingram2015a}). This misalignment introduces a level of
asymmetry not captured by our simple parameterisation of the disc
illumination profile, since in our parameterisation the disc
illumination profile is asymmetric throughout the precession
cycle. For a misaligned system, in contrast, the illumination profile
will be maximally asymmetric when the flow and disc are maximally
misaligned, and will be axisymmetric when the flow and disc align. In
other words, $A_1$ and $A_2$ would depend on QPO phase, becoming zero
once per precession cycle, rather than remaining constant as we assume
here. Nonetheless, it is clearly sensible to fit using the simplest
possible model before introducing further complexity. 

If the angle between the disc and flow is indeed changing during a
precession cycle, this will drive changes in the reflection
fraction. This is what we see in Fig. \ref{fig:paras} with $3.5\sigma$
significance. In any case, this is indicative of systematic changes of
the accretion geometry during a QPO cycle and provides yet more
confirmation of the geometric origin of Type C QPOs. In the precession 
model, this implies that the flow aligns with the disc at a QPO phase
of $\sim 0.25$ cycles when the reflection fraction dips.

In order to reproduce the observed QPO amplitude, the precession
model requires $\beta \sim 10-15^\circ$ (\citealt{Veledina2013};
\citealt{Ingram2015a}). Since here we measure the angle between our
line-of-sight and the disc spin axis, $i$, and \cite{Steiner2012} used
proper motion of the jet lobes to measure the angle between our
line-of-sight and the jet, $\theta$, we can place some constraints on
the misalignment angle $\beta$ (assuming the jet can be used as a
proxy for the BH spin axis). Even if we know $i$ and $\theta$ to
perfect precession, there is some unknown azimuthal angle,
$\Phi$. Defining $\Phi$ on the disc plane following \cite{Ingram2015a}
and \cite*{Veledina2013}, the angles $\theta$ and $i$ are related as
\begin{equation}
\cos \theta = \sin i \sin\beta \cos\Phi + \cos i \cos\beta.
\label{eqn:beta}
\end{equation}
This is equation 3 in \cite{Ingram2015a} and can be most easily
derived using the coordinate system of \cite*{Veledina2013} (see their
Fig. 2). In their formalism, $\theta$ is the angle between the vectors
$\mathbf{\hat{J}_{BH}}$ and $\mathbf{\hat{o}}$. We solve the above equation for
$\beta$ assuming best fitting values of $i=70.68^\circ$ and
$\theta=75^\circ$, running through the full range of viewer azimuth
$\Phi$ (which is completely unknown). The result is plotted in the
main panel of Fig. \ref{fig:beta} (black line). We see that nearly the
full range of possible $\beta$ values are allowed. Note that $\beta=0$
corresponds to alignment between the disc and BH spin, and
$\beta=180^\circ$ corresponds to counter-alignment (see
\citealt{King2005} for a discussion on counter-alignment). We can take
this further by simulating Gaussian distributed random variables for
$i$ and $\theta$, and a uniformly distributed random variable for
$\cos\Phi$. Since the measurement error on both $i$ and $\theta$ is
$\sim 3^\circ$, we use this as the standard deviation for both of the
Gaussian distributions. The inset plot in Fig. \ref{fig:beta} shows
the resulting cumulative probability distribution function for
$\beta$. The grey dashed line shows $\beta=15^\circ$ which is
consistent with our measurements within $0.5\sigma$ (the probability
distribution peaks at $\sim 4^\circ$).

\subsection{Continuum flux}

\begin{figure}
	\includegraphics[angle=0,width=\columnwidth]{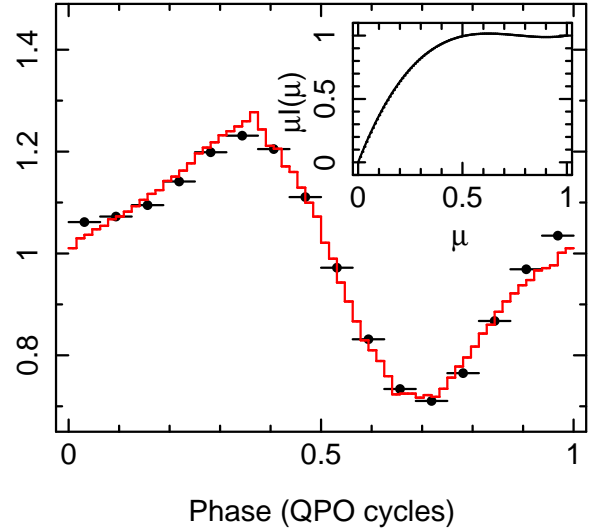}
\vspace{-9mm}
 \caption{QPO waveform for orbit 2a (black points) and for the
   precession model (red line). The observed waveform is calculated in
   units of counts $s^{-1}$ and then divided by the mean count
   rate. The waveform calculation uses the limb darkening law shown in
   the inset ($\mu$ is the cosine of the instantaneous viewing
   angle). See the text for more details.}
 \label{fig:flux}
\end{figure}

Our tomographic modeling indicates that the front and back of the disc
are preferentially illuminated by the inner flow at QPO phases $\sim
0.4$ and $\sim 0.9$ cycles. As discussed in the previous sub-section,
we also have measurements of the angles $\theta$ and $i$ and a
measurement of the disc inner radius (=flow outer radius). The bottom
panel of Fig. \ref{fig:paras} indicates that the X-ray flux peaks at $\sim 0.35$
cycles. So can the precession model reproduce this waveform in a
manner consistent with these constraints? Precession of the inner flow
will modulate the X-ray continuum flux in (at least) three ways: 1)
limb darkening, 2) changes in solid angle and 3) changes in Doppler
boosting. The limb darkening law depends on the radiative process,
which is Comptonization for the inner flow. For a stationary slab of
Comptonizing material, the observed intensity of X-ray radiation
depends on viewing angle, since photons that have undergone many
scatterings are more likely to escape at a large inclination angle
(e.g. \citealt{Sunyaev1985}; \citealt{Viironen2004}). The more face-on
we view the flow, the greater the solid angle. Without relativistic
effects, the observed flux is simply the intensity $\times$ the solid
angle. Doppler boosting has the opposite effect: the emission is
maximally boosted when the flow is viewed maximally edge-on, since this
maximizes the line-of-sight velocities. The observed flux as a
function of precession angle is then a balance between these three
considerations. Doppler boosting is most important at small radii due
to the higher rotational velocity, and solid angle effects are most
important for large radii since light bending tends to wash out solid
angle variations close to the BH (\citealt{Ingram2015a};
\citealt*{Veledina2013}).

Since our definition of QPO phase $\gamma$ is fairly arbitrary, we
must define a further parameter to tie $\gamma$ to the geometry. We
define the QPO phase such that angle between our line-of-sight and the
flow spin axis is at a minimum when $\gamma=\gamma_0$. In other words,
the flow spin axis comes the closest to pointing at the observer when
$\gamma=\gamma_0$, corresponding to a maximum in the observed solid
angle. We use the code described in \cite{Ingram2015a} to
calculate the flux as a function of $\gamma$ (i.e. the QPO waveform),
fixing $i=70.68^\circ$, $\theta=75^\circ$ and flow outer radius
$=31.47~R_g$. We leave $\gamma_0$ as a free parameter, in order to
compare the model waveform with the observed waveform. The value of
$\gamma_0$ that best reproduces the observed waveform therefore tells
about when in the QPO cycle the flow spin axis is predicted to be
maximally facing us. The code takes all relativistic effects into
account, and also includes obscuration of the flow by the disc. We
take the flow to be a torus with scale height $h/r=0.1$ (see
\citealt{Ingram2015a} for details of the flow geometry). We
parameterise the limb darkening law as
\begin{equation}
I(\mu) \propto b_0 - I_0 + b_1 \mu + b_2 \mu^2.
\label{eqn:limb}
\end{equation}
Here, $\mu$ is the cosine of the angle between the observer's
line-of-sight and the flow spin axis, and $I_0$ is set to ensure that
the minimum of $I(\mu)$ in the range $\mu=0-1$ is $I_{min}=b_0$. In
\cite{Ingram2015a} $b_0$, $b_1$ and $b_2$ were set to reproduce the
Compton scattering limb darkening law for optical depth
$\tau=1$ derived by \cite{Sunyaev1985}. Here, we leave them as free
parameters. We also leave the misalignment angle $\beta$ as a free
parameter. The remaining parameter is the inner radius of the flow, or
at least the inner radius at which the flow radiates.

We compare our waveform model to the $4-10$ keV QPO waveform of
\xmm~orbit 2a, derived using the method of IK15 (Fig \ref{fig:flux}). We perform a
least-squared fit, but note that correlated errors between the QPO
phases mean that $\chi^2$ does not give a reliable indication of
goodness of fit (see Section \ref{sec:method}). We simply use this
fitting as an exercise to try and roughly match the data. For our
`best fit' model, we set $b_0 \approx 0.3$, $b_1 \approx -1.75$ and
$b_2 \approx 0.76$, which gives the limb darkening law shown in the
inset of Fig. \ref{fig:flux}. This roughly matches the limb darkening
law expected for Compton scattering with an optical depth 
$\tau\approx 0.5$ (see Fig. 7a of \citealt{Sunyaev1985}). We set the
inner flow radius to $11~R_g$. This is fairly large, being outside of
the ISCO even for a maximally retrograde BH, but we find that
the amplitude of the waveform is sensitive to the difference between
flow outer and inner radii. This is because of the balance between
Doppler boosting and solid angle effects: the flux from a very small
radius is out of phase with the flux from a very large radius, since
the former is dominated by Doppler boosting and the latter is
dominated by solid angle variations. The amplitude of the waveform is
therefore damped by destructive interference between different
radii. It is not unreasonable for a misaligned flow to have a fairly
large inner radius, since it has been shown in General Relativistic
magneto hydrodynamic simulations that torques from the frame dragging
effect can create plunging streams at the so-called bending wave
radius, truncating the flow outside of the ISCO
(\citealt{Fragile2007}; \citealt*{Ingram2009};
\citealt{Fragile2009a}). The `best fit' misalignment angle is $\beta
\approx 11^\circ$, which is compatible with the measurements presented
in the previous sub-section.

Finally, we find a `best fit' value of $\gamma_0\approx 0.17$
cycles. This means that the flow spin axis maximally faces us at a QPO
phase of $\gamma=0.17$ cycles, and maximally faces away from us at
$\gamma=\gamma_0+0.5=0.67$ cycles. From Fig. \ref{fig:paras}, we see
that this roughly corresponds with the two maxima in line
energy. Therefore, combining tomographic modelling with the
waveform modelling implies that the flow appears to the observer to
shine preferentially on the left and right of the disc when it is
maximally facing us. This is the opposite to what I16
suggested. There, the suggestion was that the front and back of the
flow illuminate the disc such that, when the flow is facing us, it
illuminates the front and back of the disc as we see it. Whether or
not this is credible should be tested with more sophisticated
calculations. For the values we used for $\beta$, $i$ and $\theta$, it
can be derived from equation \ref{eqn:beta} that $\Phi \approx
110^\circ$. Further taking into account the fitted value of
$\gamma_0$, indicates that the flow aligns with the disc at a QPO
phase of $\gamma \approx 0.35$ cycles. The QPO phase with the lowest
observed reflection fraction gives an independant estimate for the
alignment phase. We see in Fig. \ref{fig:paras} that the minimum in
reflection fraction occurs at $\gamma \sim 0.25$ cycles, which
disagrees somewhat with the $\sim 0.35$ cycles derived from waveform
fitting.

Nonetheless, it is encouraging that we can achieve a reasonable match
to the observed QPO waveform, given the relative simplicity of both
our tomographic and waveform models. A modulation mechanism our
waveform model does not take into account is variation of seed
photons. As the misalignment angle between the disc and flow changes
over a precession cycle, the flow sees a varying luminosity of disc
photons. This will introduce a modulation into the intrinsic
luminosity of the flow ({\.Z}ycki, Done \& Ingram in prep). This
oscillation of the misalignment angle will also drive spectral
pivoting as the disc cooling changes, in addition to the
aforementioned changes in reflection fraction.\footnote{Spectral
  pivoting can additionally result from observing through different
  optical depths as the flow precesses.} We do observe a modulation in
$\Gamma$ for our best fit tomographic model (see
Fig. \ref{fig:paras}), but this is not statistically significant. It
is also likely that the optical depth, and therefore the limb
darkening law, is a function of radius (\citealt{Axelsson2014}), which
will complicate the picture further. It is very hard to see how
alternative mechanisms for the iron line centroid modulation, such as
oscillations in the disc inner radius, ionisation parameter or radial
emissivity, can be compatible with the observed QPO waveform.

We calculate the Lense-Thirring precession frequency for a flow with a
flat surface density profile extending from $11~R_g$ to $31.67~R_g$
(see equation 1 in \citealt{Ingram2012}, where $f_{LT}$ is given in
equation 3 of \citealt{Ingram2014}), and our BH spin value of
$a=0.21$. The mass of H 1743--322 is unknown, but the mass
distribution function for Galactic BHs peaks at $\sim 6.3M_\odot$
(\citealt{Ozel2010}; \citealt{Farr2011}). Using this value for mass
gives a precession frequency of $0.25$ Hz, which matches the observed
QPO frequency well. If we instead use $M=9.3 M_\odot$,
consistent with the estimate of $M \gtrsim 9.29 M_\odot$ obtained by
\cite{Ingram2014} using high frequency QPOs,  the precession frequency
becomes $0.17$ Hz.

\subsection{Biases in the time-averaged line profile}

\begin{figure}
	\includegraphics[angle=0,width=\columnwidth]{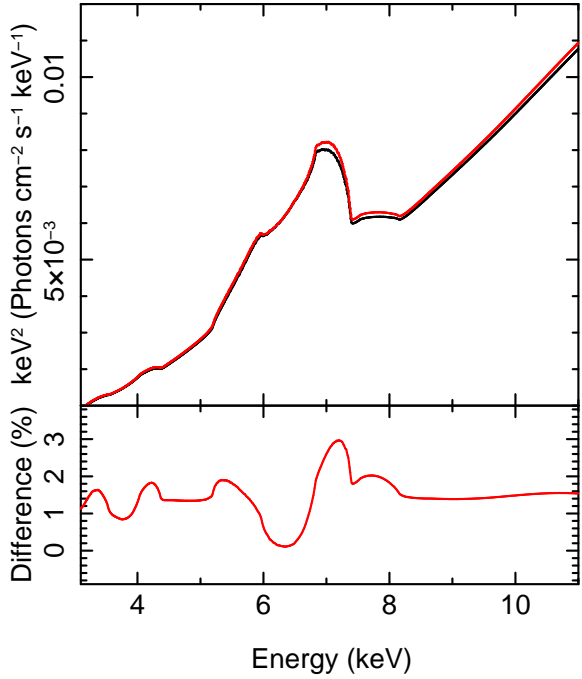}
\vspace{-9mm}
 \caption{\textit{Top:} Phase-averaged reflection spectrum,
   zoomed in on the iron line, for our best fit model (black) and also
   calculated by setting all the parameters to their phase-averaged
   values (red). A small bias is created by non-linear
   variability. \textit{Bottom:} The percentage difference
     between the two spectra (red line minus the black line, then
     divided by the black line and multiplied by $100\%$). We see an
     approximately constant $\sim 1\%$ offset, with features around
     the iron line on the $\sim 2\%$ level.}
 \label{fig:bias}
\end{figure}

For our best fitting model, the spectrum is varying with QPO phase in
a non-linear fashion. This means that the phase-averaged spectrum is
not exactly equal to a spectrum computed using the phase-averaged
parameter values. Since time-averaged spectral modelling implicitly
makes the assumption that the affects of non-linear spectral
variability are negligible, this may lead to biases. In order to
investigate these biases, we can first compare our fits to the
time-averaged spectrum in Section \ref{fig:cal} with our best fitting
tomographic model (see Tables \ref{tab:calfit} and
\ref{tab:bestfit}). We see that the tomographic modelling yields a
slightly smaller truncation radius and a slightly higher inclination,
although they are consistent within errors. The fact that these sets
of parameters are consistent with one another implies that biases due
to non-linear spectral variability (which are automatically accounted
for in our tomographic modelling but ignored for the time-averaged
spectral fits), are small.

In Fig. \ref{fig:bias}, we assess the importance of non-linear affects
more directly. The black line shows the phase-averaged reflection
spectrum corresponding to our best-fit parameters. Here, we have
calculated the spectrum for the full range of QPO phases, and
taken the mean. For the red line, we take our best fitting
model, set $A_1=A_2=0$,
$\Gamma(\gamma)=\Gamma_0$, $f(\gamma)=f_0$ (therefore removing all
non-linear variability) and calculate the phase-averaged spectrum. We
see that this slightly over-predicts the size of the blue horn, but is
a $\sim 2\%$ effect (see bottom panel). Therefore, we
conclude that the bias is small and likely does not introduce
significant systematic errors into time-averaged spectral fits, at
least for these data. We do stress, however, that considering
variability properties in addition to the time-averaged spectrum is
always advantageous since it uses more information.

\subsection{Assumptions}

We have developed the first physical model for QPO phase-resolved
spectroscopy. There are a number of improvements that could be made to
our physical assumptions in future. The model we use for the
reflection continuum, $xillver$, is the current state-of-the-art, but
improvements are still being made. First of all, $xillver$ models the
illuminating continuum as an exponentially cut-off power-law, whereas
a sharper high energy cut-off is associated with thermal Compton
up-scattering (\citealt*{Zdziarski1996}; \citealt{Fabian2015}). Also,
the disc is assumed to be a constant density slab. Making the more
physical assumption of hydrostatic equilibrium affects the predicted
reflection spectrum, but less so in the $>4$ keV range we consider
(\citealt*{Nayakshin2000}; \citealt{Done2007a}). For this paper, we
simply parameterise the radial dependence of the irradiating flux as
$r^{-3}$. This will be true far from the BH, but not close to
the irradiating source (e.g. \citealt{Laor1991,Wilkins2012}). We have also made
the simplifying assumption that the rest-frame reflection spectrum is
the same for the whole disc, allowing us to convolve the rest-frame
spectrum with a smearing kernel. However, in reality the ionisation
parameter will depend on radius since disc irradiation depends
strongly on proximity to the continuum source. \cite{Svoboda2012}
showed that not accounting for this can lead to measurement of very
centrally peaked emissivity profiles ($\sim r^{-7}$), as is often the
case (e.g. \citealt{Wilkins2011}; \citealt{Fabian2012}). Also, light
bending means that different parts of the disc have different observed
inclination angles, which makes a difference to the spectrum because
of the limb darkening law of reflected emission
(\citealt{Svoboda2009}; \citealt{Garcia2014}). Finally, our assumed
azimuthal emissivity profile is rather simplistic, but this allows us
to define a generic model to compare with the data. This can be
calibrated against more involved theoretical modeling in future.

\section{Conclusions}
\label{sec:conclusions}

We have developed the first physical model for QPO phase-resolved
spectroscopy and fit it to data from the BH binary system H
1743--322. We find that the reflection fraction varies systematically
with QPO phase 
($3.52\sigma$), adding to the now formidable body of evidence in favour of a
geometric origin of Type C QPOs. Our model mimics the asymmetric
illumination pattern, rotating about the disc surface, that would be
produced by a precessing inner flow with a simple analytic
parameterisation. It provides a good description of the observed
shifts in the iron line energy and is preferred over a null-hypothesis
of axisymmetric illumination with $2.40\sigma$ significance. More
data is therefore needed if a direct $3\sigma$ detection of asymmetric
disc illumination is to be achieved. We consider alternative
axisymmetric models, but none of them adequately describe the
data. Our results, alongside the results of I16, provide strong
evidence that Type C QPOs are driven by precession. We note that
precession of the disc rather than the flow is also possible. We
expand upon our results by modelling the continuum flux as a function
of QPO phase with a precessing inner flow model
(Fig. \ref{fig:flux}), and find we can match the observed QPO waveform
for a specific geometry in which the flow spin axis faces us at a QPO
phase of $\sim 0.2$ cycles. Since this roughly coincides with a
maximum in iron line centroid energy, this implies that the flow
preferentially illuminates the left and right hand sides of the disc
when it maximally faces us. This geometry can be tested with direct
modelling of the illumination profile from a precessing flow in
future, together with more sophisticated continuum flux waveform
modelling. Tomographic modelling of QPOs is a powerful new
technique. The next step is to apply the technique to more data in
order to track changes in accretion geometry of a source throughout an
outburst.

\section*{Acknowledgements}

We thank Chris Done, Lucy Heil and Magnus Axelsson for useful
conversations. A I acknowledges support from the Netherlands
Organization for Scientific Research (NWO) Veni Fellowship, grant
number 639.041.437. M J M appreciates support via an STFC Ernest
Rutherford Fellowship. D A acknowledges support from the
Royal Society. We thank the anonymous referee for constructive
comments that improved the paper.



\bibliographystyle{/Users/adamingram/Dropbox/bibmaster/mn2e}
\bibliography{/Users/adamingram/Dropbox/bibmaster/biblio}

%
%
%
%

\bsp	
\label{lastpage}
\end{document}